\documentclass{jfm}
\usepackage{graphicx,overpic,tikz}
\usepackage{epstopdf, epsfig}
\usepackage{amsfonts,amsbsy,ifthen,rotate,amsgen,amssymb}
\usepackage{amsmath}
\usepackage[english]{babel}
\usepackage{mathrsfs}
\usepackage[utf8]{inputenc} 
\usepackage[T1]{fontenc}
\usepackage{mathenv}
\usepackage{psfrag}
\usepackage{amsmath}
\usepackage{mathenv}
\usepackage{epsfig}
\usepackage{float}
\usepackage{natbib}
\usepackage{layout}
\usepackage{float}
\usepackage{pstricks}

\newcommand*\circled[1]{\tikz[baseline=(char.base)]{
\node[shape=circle,draw,inner sep=1](char) {#1};}}

\renewcommand{\vec}[1]{\ensuremath{\mbox{\boldmath$#1$}}}

\shorttitle{Sphere settling in a stratified fluid}
\title{Inertial drag on a sphere settling in a stratified  fluid}
\author{R. Mehaddi$^1$, F. Candelier$^2$ and B. Mehlig$^3$}
\affiliation{
$^1$ Universit\'e de Lorraine, UMR CNRS 7563, LEMTA (Laboratoire d'\'Energ\'etique et  M\'ecanique Th\'eorique et Appliqu\'e ),  F-54500   Vandoeuvre-Les-Nancy, France\\
$^2$Universit\'e d'Aix-Marseille,  UMR CNRS 7343, IUSTI  (Institut Universitaire des Syst\`emes Thermiques et Industriels)
F-13013 Marseille, France\\
$^3$ Department of Physics, Gothenburg University,
 SE-41296 Gothenburg, Sweden\\
}

\begin{document}
\maketitle
\begin{abstract}
We compute the drag force on a sphere settling slowly in a quiescent, linearly stratified fluid. Stratification can significantly enhance the drag experienced by the settling particle.  The magnitude of this effect depends on whether fluid-density transport around the settling particle is due to diffusion, to advection by the disturbance flow caused by the particle, or due to both. It therefore matters how efficiently the fluid disturbance is convected away from the particle by fluid-inertial terms.  When these terms dominate,  the Oseen drag force must be recovered. We compute by perturbation theory how the Oseen drag is modified by diffusion and stratification.  Our results are in good agreement with recent direct-numerical simulation studies of the problem at small Reynolds numbers and large (but not too large) Froude numbers.
\end{abstract}

\begin{keywords}

\end{keywords}
\section{Introduction}

The settling of small solid particles in either gaseous or liquid  flows with density stratification is a topic of great interest in fluid dynamics. Such multi-phase flows are widely encountered in Nature, in lakes or in the oceans, for example, where density stratification is due to either salt-concentration or  temperature gradients \citep{Gua12}. 
More generally, density-stratified fluids occur frequently in industrial processes that involve heated fluids \citep{linden99}, or the mixing of fluids of different densities \citep{turner79}. 

Here we compute the drag force on a sphere settling slowly in a quiescent linearly stratified fluid. The density gradient points in the direction of gravity, so that the heavier fluid is at the bottom.  
Stratification can significantly slow down the settling particle by enhancing the drag it experiences \citep{Yick09}. The reason is that buoyancy differences due to the stratification tend to prevent the vertical motion of the fluid that the particle stirs up as it settles.
As a consequence, the disturbance flow remains confined around the particle \citep[][]{Ardekani10}. How much the particle is slowed down  depends on the mechanisms that govern the dynamics of the fluid density:  diffusion of concentration or temperature,  or their advection  by the disturbance flow, or a combination of diffusion and advection. 

Which of these mechanisms is most important depends on the physical system in question. In salt water, for example,  the diffusion coefficient of salt  is much smaller than the kinematic viscosity of the fluid.  Therefore salt water is often considered non diffusive. But when temperature comes into play this may not be a good approximation, because  the  diffusion coefficient of temperature in water is roughly of the same order as  the kinematic viscosity. 
This is even more important in gases where the temperature diffusion coefficient may exceed  the kinematic viscosity of the fluid \citep{salazar03}.

The nature of the disturbance flow caused by the settling particle depends on how efficiently the fluid disturbance  is convected away. This is an inertial effect. So stratification, diffusion, and convective fluid inertia compete to determine the drag force on the particle. When the convective fluid-inertia terms dominate -- so that stratification and diffusion do not matter --  the Oseen drag force must be recovered. The question is how the Oseen drag on the settling sphere is modified by diffusion and stratification. 

The importance of convective fluid inertia is measured by the particle Reynolds number, ${\rm Re}$. The relative importance of advection and diffusion is characterised by
the P\'eclet number ${\rm Pe}$. The importance of stratification is often quantified by the viscous Richardson number ${\rm Ri}$, the ratio of buoyancy and viscous forces \citep{Yick09}.
Recent direct-numerical simulation studies of the problem \citep{Yick09,Magnaudet2018} explored how the drag depends on the importance of diffusivity versus advection, and upon the degree of density stratification. 
Our goal is to explain their results by perturbation theory, assuming that both ${\rm Re}$ and ${\rm Ri}$ are small but finite. 

\cite{Chadwick1,Chadwick2} analysed this question, but  for a sphere moving horizontally in a quiescent non-diffusive stratified fluid, along surfaces of constant fluid density. Here we study the settling problem, where the particle settles vertically along the fluid-density gradient, so that it crosses the surfaces of constant density. The two problems are quite different: when the particle moves horizontally, the  streamlines of the flow  tend to encircle the sphere in the horizontal plane.
When the sphere moves vertically,  by contrast, light fluid is pushed down into regions of larger fluid density, giving rise to complex disturbance-flow patterns \citep[][]{Ardekani10}. 

Neglecting effects of convective fluid inertia, the difference between horizontal and vertical motion was compared earlier. When
density transport is entirely diffusive,  the additional drag due to stratification is five times larger in the vertical than in the horizontal direction \citep{Candelier2014}. When density advection dominates,
the vertical drag is seven times larger than the horizontal one  \citep{Zvirin}.

Despite these qualitative and quantitative physical differences, the horizontal and vertical problems share an important mathematical property: regular perturbation expansions fail
to describe the effects of convective fluid inertia and buoyancy due to stratification even if these perturbations are weak. Therefore so-called
\lq singular-perturbation\rq{} methods are required to solve the problem. We use the standard method of asymptotic matching \citep[][]{Saffman}, where inner and outer solutions of the disturbance problem are matched, describing the disturbance flow close to and far from the particle. 

We parameterise the effect of convective inertia and stratification in terms of length scales: the particle radius $a$, the Oseen length $\ell_o = a/{\rm Re}$, and the stratification length $\ell_s = (\nu \kappa/N^2)^{1/4}$. Here $\nu$ is the kinematic fluid viscosity, $\kappa$ is the diffusivity, and $N$ is the Brunt-\!Vaisala frequency. 
The importance of diffusivity is characterised by the Prandtl number ${\rm Pr} = {\rm Pe}/{\rm Re}$ \citep{Candelier2014,Doostmohammadi2014,Magnaudet2018}.
We obtain a uniformly valid perturbation theory 
to first order in $\epsilon = a/\ell_s$ and show that analysing the results in terms of  the dimensionless parameter $\ell_s/\ell_o$ reveals three distinct regimes where density diffusion, density advection, and convective fluid inertia dominate, respectively. Fluid inertia begins to matter when 
when $\ell_s/\ell_o$ is of the order of or larger than ${\rm Pr}^{-1/4}$. At small particle Reynolds number this condition
corresponds to ${\rm Fr}\sim{\rm Re}^{-1}$, 
where ${\rm Fr} = \sqrt{{\rm Re}/{\rm Ri}}$ is the Froude number. This condition
is met  in recent direct numerical simulations (DNS) of the problem \citep{Yick09,Magnaudet2018}, and our results are in good agreement
with the simulations at small ${\rm Re}$ and at ${\rm Fr}\sim 10$.
When the Fr is much larger, then finite-size effects in the DNS give rise to deviations from our theory for the 
unbounded system.
Small values of ${\rm Fr}$ correspond to large values of $\epsilon$. Here
the theory fails because it requires $\epsilon$ to be small.

\section{Formulation of the problem}
We consider a spherical particle of radius $a$ and of material density $\rho_p$ settling  with velocity $\vec{u}$ in a quiescent stratified fluid. The diffusivity of the stratifying agent (salt or temperature) is denoted by $\kappa$, and the kinematic viscosity of the fluid is denoted by $\nu$. The ambient density of the fluid is assumed
to vary linearly with height $z$
\begin{equation}
\rho_{0} = \rho_{\infty}-\gamma z\,,
\end{equation}
where $\gamma$ is the density gradient, and $\rho_{\infty}$ is a reference density. 
We assume that quadratic combinations of the density and pressure disturbances are negligible, and that $\gamma z/\rho_\infty \ll1$ in the region of interest.
This allows us to ignore density gradients except when multiplied by
the gravitational acceleration 
\citep{gray76}. This \lq Boussinesq\rq{} approximation was used in the DNS of the problem by \cite{Yick09} and \cite{Magnaudet2018} that we compare with in Section \S 5, see also \citep{Doostmohammadi2014}. 
When a particle settles in a stratified fluid, it experiences a time-dependent buoyancy force, because the unperturbed density $\rho_0$ varies as a function of height $z$. Under the Boussinesq approximation this variation is negligible, so that the particle  reaches a quasi-steady settling velocity. We consider this steady limit.
In a quiescent fluid, the velocity disturbance $\vec{w}$ is simply the flow produced by  the particle.  Its motion modifies the local density and pressure, and  we define density and pressure disturbances as  $\rho' = \rho-\rho_0$ and $p'=p-p_0$. Here $p_0$ is the hydrostatic pressure.  These disturbances are determined by:
\begin{subequations}
\label{eq:ns}
\begin{align}
&\mbox{Re}\big[ \left(\vec{w} \cdot \vec{\nabla}\right)\vec{w}- \left(\vec{u} \cdot \vec{\nabla}\right)\vec{w} \big] +\mbox{Ri}\:\rho'\hat{\bf e}_3 =-\vec{\nabla} p'+ \vec{\Delta}\vec{w} \quad\mbox{and}\quad \vec{\nabla}\cdot \vec{w}=0\,,\label{quantite de mouvement3} \\
&\mbox{Pe}\big[ \left(\vec{w} \cdot \vec{\nabla}\right)\rho' -  \left(\vec{u}\cdot \vec{\nabla}\right)\rho' - \vec{w} \cdot \hat{\bf e}_3 \big] = \vec{\Delta}\rho'\:.\label{advection_diffusion3}\\
\label{eq:bc}
& \vec{w}=\vec {u}\,,\left.\partial_r\rho'\right\vert_{r=1} = \cos\theta
 \quad\mbox{at}\quad r=1 \quad  \mbox{and}\quad \vec{w}\rightarrow \vec{0}\,,\rho'\to 0\quad\mbox{as}\quad \: r\rightarrow \infty\,.
\end{align}
\end{subequations}
Here  $\theta$ is the angle between the outward unit normal $\vec{n}$
of the sphere and the vertical direction $\hat {\bf e}_3$.
The boundary condition for $\rho'$ on the surface of the particle
is derived from the surface condition $\vec{\nabla} \rho \cdot \vec{n} = 0$.
This means that the particle surface is impermeable. 

We  de-dimensionalised the problem in the usual fashion \citep{Alias17},
using the particle radius $a$ for lengths, the terminal Stokes velocity $u_t=[9a^2/(2\nu)](\rho_p/\rho_\infty-1)g$ for the fluid velocity (where $g$ is the gravitational acceleration), $\rho_\infty \nu u_t/a$ for the pressure, and $\gamma a$ for the density.  
The dimensionless parameters  in Eqs.~(\ref{quantite de mouvement3}) and (\ref{advection_diffusion3})  are the particle Reynolds number, the P\'eclet number, and the Richardson number:
\begin{equation}
\mbox{Re} = {a u_t}/{\nu} \:, \quad \mbox{Pe} = {a u_t}/{\kappa} \,, \quad \mbox{and} \quad \mbox{Ri} = {a^3N^2}/({u_t \nu})\:.
\end{equation}
Here  $N$ is the Brunt-\!Vaisala frequency
\begin{equation}
N={\sqrt{g\,\gamma/\rho_{\infty}}}\:,
\end{equation}
the  frequency at which a perturbation describing  a vertically displaced parcel of fluid oscillates within a statically stable environment  \citep{Mowbray1967}.

In this paper we obtain the drag force on the sphere  assuming  that convective fluid inertia and density stratification matter, but that they are weak enough so that their effects can be treated in perturbation theory ($0 < {\rm Re}\ll1$ and $0 < {\rm Ri}\ll 1$). 

\section{Earlier results for ${\rm Re}=0$}
For  ${\rm Re}=0$ the drag on a sphere settling  in a stratified fluid  was studied  theoretically by \cite{Zvirin}  and \cite{Candelier2014}. These authors made different assumptions concerning the relative importance of advection and diffusion in Eq.~(\ref{advection_diffusion3}).
\cite{Zvirin} assumed that advection is more important than diffusion. 
When advection dominates, 
the density disturbance $\rho'$ scales as $z/r$ near the particle \citep{Chadwick1}, in the \lq inner region\rq{} of the problem.  As a consequence, the buoyancy term in Eq. (\ref{quantite de mouvement3}) balances the viscous Laplacian term at  \begin{equation}
r \sim {\mbox{Ri}^{-1/3}}\:. 
\label{r_matching_Ri}
\end{equation}
At this distance  inner and outer solutions of the disturbance problem must be matched. This implies that  advection is more important than diffusion in Eq.~(\ref{advection_diffusion3}) if  ${\rm Pe}> \mbox{Ri}^{1/3}$. Second, at $r \sim {\mbox{Ri}^{-1/3}}$ the dominant convective inertial term in Eq. (\ref{quantite de mouvement3}) is estimated as $\mbox{Re}  (\vec{u} \cdot \vec{\nabla}) \vec{w}^{(0)} \sim  \mbox{Re}\: \mbox{Ri}^{2/3}$.
So convective inertial terms are negligible if
 $\mbox{Re} \ll \mbox{Ri}^{1/3}$. 
Under  these conditions,
\begin{equation}
\label{eq:zvirin_conditions}
\mbox{Pe} >\mbox{Ri}^{1/3} \quad\mbox{and}\quad \mbox{Re} \ll \mbox{Ri}^{1/3}\,,
\end{equation}
\cite{Zvirin} derived
the following expression for the drag force
\begin{eqnarray}
f_3 = - 6\pi u_3 \big[1+B\big({\mbox{Ri}^{1/3}}/{\mbox{Pe}}\big)\:\mbox{Ri}^{1/3} \big]\:. \label{zvirin_general}
\end{eqnarray}
Here $B(\cdot)$ is a function given in integral form. 
 In the limit of a non-diffusive fluid,  ${\rm Pe}\to\infty$, 
 the above expression simplifies to:
 \begin{eqnarray}
 f_3 = - 6\pi u_3 (1+1.060 \:\mbox{Ri}^{1/3})\:. \label{zvirin}
 \end{eqnarray}
Now consider the opposite limit, where the diffusive
term in Eq. (\ref{advection_diffusion3})  dominates over the advective term.
In this case \cite{Candelier2014}
showed that the spatial dependence of  the disturbance density $\rho'$ 
is of the form $\rho' \sim \mbox{Pe}\: r$ in the inner region, so that 
 the buoyancy term in Eq. (\ref{quantite de mouvement3}) balances the Laplacian viscous
  term at 
 \begin{equation}
\label{eq:epsilon}
r \sim{\epsilon}^{-1} \quad \mbox{with} \quad \epsilon ={a}/{\ell_s}\:.
\end{equation}
Here $\ell_s$ is the stratification length 
\citep{Ardekani10} 
\begin{equation}
\ell_s=\left({\nu \kappa}/{N^2}\right)^{1/4}\:. \label{ardekani}
\end{equation}
It characterises the effect of stratification on the particle dynamics.
Under the condition
\begin{equation}
 {\rm Pe}\ll \epsilon\ll1
\end{equation}
\cite{Candelier2014} found
\begin{eqnarray}
f_3 = - 6\pi u_3 (1+ 0.6621 \epsilon )\:. \label{candelier}
\end{eqnarray}
Using $\mbox{Ri} ={\epsilon^4}/{\mbox{Pe}}$,  we see that
the condition ${\rm Pe}\ll \epsilon$ corresponds to ${\rm Pe}\ll {\rm Ri}^{1/3}$.
Comparing with the condition (\ref{eq:zvirin_conditions}) it seems that the results (\ref{zvirin}) and (\ref{candelier}) apply in the opposite limits of large and small P\'eclet numbers. Below we show, however, that
the two approaches are in fact equivalent, although they seem to apply in distinct limits.

\section{Method}
We consider the same problem as
 \cite{Candelier2014}, but we do not neglect the fluid-inertia terms and 
 the effect of advection of the fluid density by the disturbance flow. 
The relative importance of stratification and inertial effects is determined
by the magnitude of the length scales $\ell_s$ and $\ell_o$ in relation
to the particle size $a$. Therefore we use $\epsilon = a/\ell_s$ [Eq.~(\ref{eq:epsilon})]
and $\ell_s/\ell_o$ as dimensionless parameters. The third parameter is the Prandtl  number.
In summary, we solve Eqs.~(\ref{eq:ns}) to first order in the parameter $\epsilon$ using the method of asymptotic matching \citep{Saffman}. Inner and outer solutions of the disturbance problem are matched at $r \sim \epsilon^{-1}$ in the limit 
\begin{equation}
\label{eq:lim}
\epsilon \ll 1 \quad\mbox{with  $\ell_s/\ell_o$ and $\mbox{Pr}$ arbitrary but fixed.}
\end{equation}
In this way we obtain  an expression for drag force that is valid regardless of whether diffusion or advection dominates: our solution is valid in both limits considered by \cite{Candelier2014} and \cite{Zvirin}, as well as uniformly in between. 

Previous arguments, summarised in \S 3, appeal to different behaviours of the density disturbance to show that the non-linear convective terms $\mbox{Re} (\vec{w}^{(0)} \cdot \vec{\nabla}) \vec{w}^{(0)}$ and $\mbox{Pe}\,(\vec{w}^{(0)}\cdot \vec{\nabla})\rho'$ in Eq.~(\ref{eq:ns}) can be disregarded. A weakness of these arguments  is that the limits of large and small Pe are considered separately.  This is not necessary in our formulation.

A general  property of the method of asymptotic matching is that it is the magnitude of the different terms in the matching region that matters: all terms
that are sub-leading in this region can be entirely neglected.
When ${\rm Re}$ and ${\rm Ri}$ are small, the disturbance velocity close to the particle is well approximated by the Stokes solution  $\vec{w}^{(0)} \sim  {1}/{r}$. Assuming this dependence we can estimate the magnitude
of the non-linear convective term $\mbox{Re} (\vec{w}^{(0)} \cdot \vec{\nabla}) \vec{w}^{(0)}$ in the matching region.
Setting $r\sim \epsilon^{-1}$ we conclude that $\mbox{Re} (\vec{w}^{(0)} \cdot \vec{\nabla}) \vec{w}^{(0)}$  is small in this region compared with all other terms in Eq.~(\ref{quantite de mouvement3}), when $\epsilon$ is small.
The orders of magnitude in Eq.~(\ref{advection_diffusion3}) are more difficult to determine because the $\vec{r}$-dependence of the density disturbance is not known unless Pe is either small \citep{Candelier2014} or large \citep{Zvirin}. However, since  $\vec{w}^{(0)} \sim \epsilon$ in the matching region, we can conclude that the non-linear term $\mbox{Pe}\,(\vec{w}^{(0)} \cdot \vec{\nabla})\rho'$
is  negligible compared with $\mbox{Pe}\,(\vec{u} \cdot \vec{\nabla})\rho'$.  
As a result, Eqs.~(\ref{eq:ns}) take the form:
\begin{subequations}
\label{eq:form_prob}
\begin{align}
& - \epsilon\: \frac{\ell_s}{\ell_o}  \left(\vec{u} \cdot \vec{\nabla}\right)\vec{w}  =-\vec{\nabla} p'-\epsilon^4\tilde{\rho}\hat{\bf e}_3 + \vec{\Delta}\vec{w} \quad\mbox{and}\quad \vec{\nabla}\cdot \vec{w}=0\:,
\label{quantite de mouvement4} \\
&-\epsilon\: \mbox{Pr} \frac{\ell_s}{\ell_o}(\vec{u} \cdot \vec{\nabla})\tilde{\rho} - \vec{w}\cdot\hat{\bf e}_3=\vec{\Delta}\tilde{\rho}\:, \label{advection_diffusion4}
\end{align}
\end{subequations}
with boundary conditions corresponding to (\ref{eq:bc}), and   $\rho'= \: \mbox{Pe}\, \tilde{\rho}$. 
The inner solution of Eqs.~(\ref{eq:form_prob}) is obtained by a regular perturbation expansion  in $\epsilon$. To obtain the outer solution one replaces
the boundary condition on the particle surface by a singular source term \citep{Saffman},  of the form $6\pi \vec{u}\,\delta(\vec{r})$.
Since the non-linear convective terms are negligible, Eq.~(\ref{eq:form_prob}) is linear, so that the outer solution 
can  be obtained by Fourier transform, for arbitrary values of $\epsilon$. We define:
\begin{equation}
\hat{f}(\vec{k})  =  \int\!\mathrm{d}\vec{x}\, f(\vec{x}){\rm e}^{-i \small \vec{k}\cdot \vec{x}} \quad \mbox{and} \quad f(\vec{x}) = \int
\!\!\frac{\mathrm{d}\vec{k}}{\small (2\pi)^3} \hat{f}(\vec{k}){\rm e}^{i \small \vec{k}\cdot \vec{x}}  \:.
\end{equation}
We expand the Fourier transform $\hat{\vec{w}}_{\mbox{\scriptsize out}}(\vec{k})$ of the outer solution in $\epsilon$, in terms of generalised functions  \citep{Candelier2013,Meibohm16}:
\begin{equation}
\label{eq:exp}
\hat{\vec{w}}_{\mbox{\scriptsize out}} = \hat{\vec{\mathcal{T}}}^{(0)} + \epsilon \hat{\vec{\mathcal{T}}}^{(1)}  +
\epsilon^2 \hat{\vec{\mathcal{T}}}^{(2)} + \ldots .
\end{equation}
This method differs slightly from the standard approach \citep{Saffman} that formulates the outer problem in terms
of strained coordinates $\overline{\vec{r} }= \epsilon \vec{r}$. The advantage of the present approach is that it does not
refer to any particular matching length scale -- for instance the length scale at which the Laplacian is balanced by the buoyancy term in Eq.~(\ref{quantite de mouvement4}). The only requirement is that $\epsilon $ is small. For certain cases this approach is equivalent to using the reciprocal theorem to compute inertial corrections \citep{Meibohm16}.

The first two terms in the expansion (\ref{eq:exp}) are obtained from $\hat{\vec{w}}_{\mbox{\scriptsize out}} $ as:
\begin{equation}	
\hat{\vec{\mathcal{T}}}^{(0)} = \lim_{\epsilon \to 0 } \hat{\vec{w}}_{\mbox{\scriptsize out}} 
\:\:\:\: \mbox{and} \:\:\:\:
\hat{\vec{\mathcal{T}}}^{(1)} = \lim_{\epsilon \to 0 } \tfrac{1}{\epsilon} ( \hat{\vec{w}}_{\mbox{\scriptsize out}} -\hat{\vec{\mathcal{T}}}^{(0)} )\:.
\label{Eq_limit}
\end{equation}
The first term, $\hat{\vec{\mathcal{T}}}^{(0)}$, is the Fourier transform of 
the solution of the outer problem  at $\epsilon=0$. The next term in the expansion reads
\citep{Candelier2013,Meibohm16}
\begin{equation}
\hat{\vec{\mathcal{T}}}^{(1)} = \delta(\vec{k}) \, \int \!{\rm d}\vec{k}\,\big( \vec{\hat{w}_{\mbox{\scriptsize out}}}|_{\epsilon=1}-\hat{\vec{\mathcal{T}}}^{(0)} \big) \:.
\end{equation}
The functions $\hat{\vec{\mathcal{T}}}^{(0)}$ and $\hat{\vec{\mathcal{T}}}^{(1)}$ are readily transformed back
to obtain the outer solution in configuration space. 
In particular, $\vec{\mathcal{T}}^{(1)}(\vec{r})$ is found to  be  $\vec{r}$-independent.
 Since $\vec{\mathcal{T}}^{(1)}(\vec{r})$ is constant, the problem to order $\epsilon$ is equivalent to
 determining the force on a particle kept fixed in a uniform flow \citep{Saffman,Meibohm16}.
It follows that the drag force is given by 
\begin{equation}
\vec{f}=-6\pi  \left[\vec{u} + \frac{\epsilon}{8\pi^3}  \int\!{\rm d}\vec{k}\, \big( \vec{\hat{w}_{\mbox{\scriptsize out}}}|_{\epsilon=1}-\hat{\vec{\mathcal{T}}}^{(0)} \big)  \right]\:. \label{procedure}
\end{equation}
We note that the force  is determined entirely by the  solution of the outer problem,
as first shown by \cite{Saffman} for the lift force on a small sphere in a shear flow.

\section{Results}
For  $\epsilon=1$ the Fourier transforms $\hat{\vec{w}}_{\mbox{\scriptsize out}}$ and $\hat{\tilde{\rho}}_{\mbox{\scriptsize out}}$ of the outer solution read:
\begin{equation}
\left(\begin{array}{c} 
\vec{\hat{w}_{\mbox{\scriptsize out}}}\\
\hat{\tilde{\rho}}_{\mbox{\scriptsize out}}
\end{array}
\right) = - 6\pi \:  k^2 \Big[\frac{\ell_s}{\ell_o} \Big(i \vec{k} \cdot \vec{u} \Big) \mathbb{I}+\mathbb{A} \Big]^{-1} \cdot \mathbb{G} \cdot \left(\begin{array}{c} 
\vec{u}\\
0
\end{array}
\right)\:.
\end{equation}
Here  $\mathbb{I}$ is the $4\times4$ unit tensor, and
\begin{equation}
 \mathbb{A}=\!  \left( \begin{array}{cccc}
- k^2 &0    &0   &   -  \frac{k_1k_3 }{k^2}    \\ 
 0  &- k^2&0  &   -  \frac{k_2k_3 }{k^2}  \\ 
 0    &0  &- k^2  &   - \frac{(k^2-k_3^2) }{k^2}   \\ 
 0   &0  &\frac{1}{\mbox{\scriptsize Pr}}   &- \frac{k^2}{\mbox{\scriptsize  Pr} }  \\ 
\end{array} \right)\!,
\quad\mathbb{G}=\! \left( \begin{array}{cccc}
  \frac{k^2-k_1^2}{k^4} & -\frac{k_1k_2}{k^4} & -\frac{k_1k_3}{k^4} &0  \\ 
 -\frac{k_2k_1}{k^4} & \frac{k^2-k_2^2}{k^4} & -\frac{k_2k_3}{k^4}&0 \\
 -\frac{k_3k_1}{k^4} & -\frac{k_3k_2}{k^4} & \frac{k^2-k_3^2}{k^4}&0 \\ 
  0&0&0&0  \\ 
\end{array} \right) \:.
\end{equation}
We set $\vec{u}=u_3 \hat{\bf e}_3$  in Eq.~(\ref{procedure})  to find the drag force  on the settling sphere:
\begin{subequations}
\begin{align}
f_3& = - 6\pi u_3 (1+\epsilon \: M_{33} )\:, \label{force}\\
{M}_{33} &= \frac{3 }{2\pi  }\int_0^{\infty}\!\! \!\!\mbox{d}k \!\int_0^{\pi} \!\!\mbox{d}\theta \,{\frac {\sin(  \theta )^{3} \Big\{1\!-\! \Big[\mbox{Pr}\,{\left(\frac{\ell_s}{\ell_o}\right)}^{2}{k}^{2}\!+\!1 \Big]  \cos \left( \theta \right)^{2}\!-i\cos \left( \theta \right) \frac{\ell_s}{\ell_o} {k}^{3} \Big\}  }{  {\Big[ \mbox{Pr}\,{\left(\frac{\ell_s}{\ell_o}\right)}^{2}{k}^{2}\!+\!1 \Big] \cos \left( \theta \right)^{2}\!+i\frac{\ell_s}{\ell_o}{k}^{3} \left( \mbox{Pr}\!+\!1 \right) \cos \left( \theta \right)\! -\!{k}^{4}\!-\!1 }}}\,. \label{Eq_M33}
\end{align}
\end{subequations}
The imaginary part in Eq. (\ref{Eq_M33}) vanishes upon integration.

\begin{figure}
\begin{overpic}[width=13cm]{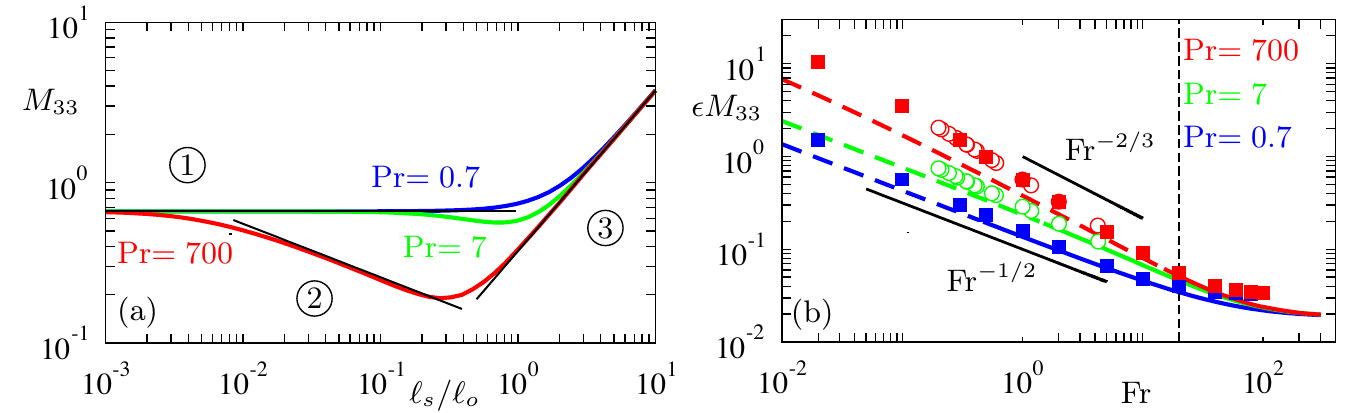}
\end{overpic}
\caption{
(a) Correction to the drag, Eq. (\ref{Eq_M33}), as a function of  $\ell_s/\ell_o$ for different $\mbox{Pr}$. Also shown are
the three different regimes in Eq.~(\ref{eq:summary}), black solid lines.
(b) Comparison between Eq. (\ref{Eq_M33}) and DNS results 
for Re$=0.05$ by \cite{Yick09} for Pr$=7$ (\textcolor{green}{\LARGE $\circ$}),
Pr$=700$ (\textcolor{red}{\LARGE $\circ$}), and by
\cite{Magnaudet2018} for Pr$=0.7$ (\textcolor{blue}{$\blacksquare$})
and Pr$=700$ (\textcolor{red}{$\blacksquare$}).
 Coloured solid lines show Eq.~(\ref{Eq_M33}) for 
$\epsilon <0.3$,  dashed lines for $\epsilon>0.3$. 
Also shown are power laws in Fr, black solid lines. The vertical dashed line corresponds to Fr=1/Re. }
\label{fig_M33}
\end{figure}
Fig. \ref{fig_M33}(a) shows how $M_{33}$ depends on the ratio $\ell_s/\ell_o$ for different values of Pr, namely, 0.7 (temperature-stratified gas), 7 (temperature-stratified water at $20^{\:\mbox{\tiny o}} C$) and 700 (salt-stratified water).  When  the ratio $\ell_s/\ell_o$ is very small,  the curves collapse onto  a horizontal line, Eq.~(\ref{candelier}). In this limit 
diffusion dominates. 
When $\ell_s/\ell_o$ reaches ${\rm Pr}^{-1}$, a second regime emerges: 
diffusion and  advection in  Eq. (\ref{advection_diffusion3}) become of the same order,
resulting in a change in the behaviour of the density disturbance from $\rho' \sim \mbox{Pe} \:r$ to $\rho' \sim z/r$. As a result, 
the curves in Fig. \ref{fig_M33} turn downwards. A further transition occurs at $\ell_s/\ell_o \sim 1/\mbox{Pr}^{1/4}$, caused by the formation of an Oseen wake behind the particle \citep{Lovalenti}.  When $\ell_s/\ell_o  \gg 1/\mbox{Pr}^{1/4}$ the curves approach  $M_{33}\approx (3/8)(\ell_s/\ell_o) $,  the Oseen correction \citep{Oseen10,Proudman57,Lovalenti}. In this regime stratification and diffusion do not matter, the settling particle experiences the fluid as if it were homogeneous.  For small Pr, only the first and third regimes are realised.
 
Eq.~(\ref{Eq_M33}) is uniformly valid in the limit (\ref{eq:lim}), 
regardless of the value of $({\ell_s}/{\ell_o}) \mbox{Pr} = ({{\mbox{Ri}^{1/3}}/\mbox{Pe}})^{-3/4}$.
It is not necessary to assume that  ${\rm Pe}\ll {\rm Ri}^{1/3}$, the expression holds also when ${\rm Pe}\gg {\rm Ri}^{1/3}$. In particular, we can see that Eq.~(\ref{Eq_M33}) reduces
to Eq.~(\ref{zvirin_general}) when convective inertia is negligible, 
by taking the limit  $\ell_s/\ell_o \to 0$  at fixed 
$({\ell_s}/{\ell_o}) \mbox{Pr}$:
\begin{equation}
\lim_{\ell_s/\ell_o \to 0} M_{33} = \frac{\mbox{3\,Ri}^{1/3}}{\pi\epsilon} 
\int_0^\infty\!\!\!\!  \mbox{d}k \int_0^{\frac{\pi}{2}}\! \!\!\mbox{d}\theta
\frac{\sin(\theta)^5\big( \sin(\theta)^2 + (\mbox{Ri}^{1/3}/\mbox{Pe})k^4\big)}{\big( \sin(\theta)^2 + (\mbox{Ri}^{1/3}/\mbox{Pe})k^4\big)^2 + \cos(\theta)^2 k^6} \:.
\label{Eq_Zvirin}
\end{equation}
This is precisely the function $B(\cdot)$ in Eq.~(\ref{zvirin_general}), Eq.~(29) in \citep{Zvirin}, computed assuming that convective inertia is negligible, and that ${\rm Pe}> {\rm Ri}^{1/3}$.
Since our solution is uniformly valid, we can conclude that Eq.~(\ref{zvirin_general})
must be valid also for ${\rm Pe}\ll {\rm Ri}^{1/3}$, well outside the region of validity stated
by \cite{Zvirin}. Closer inspection of their calculation shows that it corresponds to asymptotic matching at $r\sim {\rm Ri}^{-1/3}$ in the limit ${\rm Ri}\to 0$ keeping ${\rm Ri}^{1/3}/{\rm Pe}$ constant. The two different matching scales  $r\sim {\rm Ri}^{-1/3}$ and $r\sim\epsilon^{-1}$
are equivalent in the limits stated, because the ratio of matching scales
${\rm Ri}^{1/3}/\epsilon = ({\rm Ri}^{1/3}/{\rm Pe}) ({\rm Pe}/\epsilon)  = ({\rm Ri}^{1/3}/{\rm Pe})  [(\ell_s/\ell_o) {\rm Pr}]^{-1}$ remains constant.   
In summary, Eq.~(\ref{Eq_M33}) is a uniform approximation comprising three distinct regimes 
\begin{equation}
\label{eq:summary}
f_3\! \sim\! - 6\pi u_3 
\left\{\begin{array}{lll} 
\!\!1\!+\!0.6621 \epsilon  & \mbox{for} \: \ell_s/\ell_o \ll \mbox{Pr}^{-1} & \mbox{regime \circled{$1$} (diffusion),} \\
\!\!1\!+\!1.060 \:\mbox{Ri}^{1/3}\hspace*{-3mm}& \mbox{for}  \:\mbox{Pr}^{-1}\!\!  \ll \!\ell_s/\ell_o \!\ll \!\mbox{Pr}^{-1/4}\hspace*{-2mm}
& \mbox{regime \circled{$2$} (advection),} \\
\!\!1\!+\!\tfrac{3}{8}\:\mbox{Re}& \mbox{for} \: \ell_s/\ell_o \gg \mbox{Pr}^{-1/4} &\mbox{regime \circled{$3$} (fluid inertia)}. \\
\end{array} \right. 
\end{equation}
The different regimes are shown in Fig.~\ref{fig_M33}(a). In the limit of small Pr, the advective regime disappears, as mentioned above. 
\newpage
We now compare the full result, Eq.~(\ref{Eq_M33}), with DNS by \cite{Magnaudet2018} and \cite{Yick09}, 
at their smallest ${\rm Re}$. In these simulations, an alternative set of parameters was used: $\mbox{Re}$, $\mbox{Pr}$, and the Froude number 
\begin{equation}
\mbox{Fr}={u_t}/({a\:N})\:.
 \end{equation}
 In terms of Fr, the
dimensionless parameters $\epsilon$, $\ell_s/\ell_o$, and  $\mbox{Ri}$ are given by:
\begin{equation}
 \epsilon = \left({\mbox{Re}}/{\mbox{Fr}}\right)^{1/2}\:\mbox{Pr}^{1/4}\:, \quad 
 {\ell_s}/{\ell_o} =  {\left( \mbox{Re}\, \mbox{Fr}\right)^{1/2} }/{\mbox{Pr}^{1/4}}\,, \quad \mbox{and} \quad{\mbox{Ri}^{1/3} = {\mbox{Re}^{1/3}}/{\mbox{Fr}^{2/3}}}   \:.
\label{Eq_epsilon_Re_Fr} 
 \end{equation}
\citet{Magnaudet2018} and \cite{Yick09} computed the drag coefficient $C_{\rm D}^{\rm S}$ of the stratified system. 
In Fig.~\ref{fig_M33}(b) we plot their result for $C_{\rm D}^{\rm S}/C_{\rm D}^{\rm Stokes}-1$ versus Fr, and compare
it with our result for $\epsilon M_{33}$. Here $C_{\rm D}^{\rm Stokes} = 12/{\rm Re}$ is the Stokes drag coefficient for an unbounded system. Since  Eq.~(\ref{Eq_M33}) was obtained for small $\epsilon$, we
plot it as a solid line when $\epsilon < 0.3$, and dashed for $\epsilon > 0.3$. 
For $1 < {\rm Fr} < 10$ the data for ${\rm Pr}=0.7$ are  in the diffusive regime, where
 the correction to the drag scales as ${\rm Fr}^{-1/2}$.  For ${\rm Pr}=700$, the data approach the advection regime where the theory predicts that the drag correction scales as  ${\rm Ri}^{1/3} \propto {\rm Fr}^{-2/3}$. But this power law is not clear cut in the DNS data.
 
 When do convective fluid-inertia effects dominate? The condition  $\ell_s/\ell_o =1/\mbox{Pr}^{1/4}$ corresponds to ${\rm Fr} = 1/{\rm Re}$, independent of Prandtl number. 
For ${\rm Re}= 0.05$ -- the smallest value used in the DNS  -- this crossover occurs at ${\rm Fr}=20$, indicated by the vertical black dashed line in Fig.~\ref{fig_M33}(b).  Eq.~(\ref{Eq_M33}) allows us to determine the relative importance of convective fluid inertia
 at this value of ${\rm Fr}$. For $\mbox{Pr}=0.7$ the correction is substantial, 13.5~\%.
 For larger P\'eclet numbers the correction is smaller, 1.4\% at $\mbox{Pr}=7$, and 2.~ \% at $\mbox{Pr}=700$.
 That the correction is largest for small Pr can be inferred from Fig.~\ref{fig_M33}(a).

Fig.~\ref{fig_M33}(b) shows that the DNS  yield a larger drag coefficient than our theory when Fr is small. The likely reason is that the non-linear
convective terms matter in this regime. But also at large Fr there are deviations. These may be due to finite-size effects. 
At very large Fr  the homogeneous Oseen correction dominates, and at small Re it is quite sensitive to the size of the simulation domain. \cite{Yick09} chose an elliptical simulation domain, with a smallest size $L$ that gives $L/(2a)=40$. The domain used 
by \cite{Magnaudet2018} was spherical and larger [diameter$/(2a)=80$], but even in that case a theory for cylindrical domains \citep{happel1983} indicates that the drag correction is expected to be higher than the Oseen expression $\tfrac{3}{8}{\rm Re}$. This is consistent with Fig.~\ref{fig_M33}(b).
Finite-size effects matter less for smaller Fr, because the wake is smaller, of order $\ell_s$.

\section{Conclusions}
We calculated how convective fluid inertia modifies the drag on a sphere slowly settling in a density-stratified fluid, at small Richardson and Reynolds numbers. Plotting the results as a function of the dimensionless parameter $\ell_s/\ell_o$ reveals three distinct regimes, Eq.~(\ref{eq:summary}). In the first regime, the drag is determined by diffusion
of the disturbance density. In the second regime, advection of the disturbance density determines the drag. In the third regime, convection of the disturbance density by fluid-inertia terms dominates.  Our main result, Eq.~(\ref{Eq_M33}), is uniformly valid, independently of whether the density dynamics is diffusive or advective. This allowed us to show that a result by \cite{Zvirin} is more generally valid than the authors stated.

We compared with recent DNS at small ${\rm Re}$ and found that 
convective fluid-inertia effects matter for the largest Froude numbers simulated, but
the fluid appears still far from homogeneous for the settling particle.

The results derived in this paper were obtained in the steady limit. But when  a particle is released from above the 
water surface and plunges into the fluid with a given velocity, then unsteady effects must matter, at least initially.  
DNS of the problem \citep{Doostmohammadi2014} at Re of order unity
reveal unsteady effects that depend on the dimensionless numbers of the problem in intricate ways. 
Since finite-size effects appear to be important at large Fr and small Re, it would be of interest to take these corrections into account in the theory.

A further motivation  for taking into account unsteady effects concerns the unsteady swimming of micro-organism in stratified fluids. In oceans or in lakes the surface layers are known to shelter substantial biological activity. For very small organisms (much smaller than 1mm in size in typical ocean conditions) the dynamics of swimming microorganisms is well understood. Buoyancy \citep{FJ2008}, density or drag asymmetries of the body \citep{Roberts1970,Jonsson1989,Kes85}, and turbulence \citep{Dur13,Berglund16} determine the spatial distribution of these organisms, their encounter rates, and thus their population ecology \citep{Gua12}. For larger organisms less is known. The problem becomes considerably more difficult because inertial effects begin to matter  \citep{Wang12,Wang12b}. The method described here allows to take inertial effects into account in perturbation theory. Finally, an important problem is how fluid shears affect the dynamics of motile microorganisms. The approach described by \cite{Candelier18} makes it possible to address this question.

\acknowledgements{
We thank J. Magnaudet and J. Zhang for providing some of the numerical data discussed in \cite{Magnaudet2018}. BM was supported by  Vetenskapsr\aa{}det [grants 2013-3992 and 2017-03865], Formas [grant number 2014-585],  and by the grant \lq Bottlenecks for particle growth in turbulent aerosols\rq{} from the Knut and Alice Wallenberg Foundation, Dnr. KAW 2014.0048.}

\end{document}